\documentstyle[aps,epsf,prl,multicol]{revtex}

\begin{document}
\draft
\title{
Heat conduction in one dimensional chains
}
\author{Bambi Hu$^{[1,2]}$, Baowen Li$^{[1]}$, and Hong Zhao$^{[1,3]}$} 
\address{
$^{[1]}$ Centre for Nonlinear Studies and  Department of Physics, Hong Kong 
Baptist University, China \\
$^{[2]}$ Department of Physics, University of Houston, Houston TX 77204 USA\\
$^{[3]}$ Department of Physics, Lanzhou University, 730000 
Lanzhou, China}
\maketitle

\begin{abstract} 
We study numerically the thermal conductivity in several different one 
dimensional chains. We show that the phonon-lattice interaction 
is the main ingredient of the Fourier heat law. Our argument provides a 
rather satisfactory explanation to all existing numerical results 
concerning this problem.
\end{abstract} 
\begin{multicols}{2}

It is still an open and challenging problem to understand the macroscopic 
phenomena and their statistical properties in terms of the deterministic 
microscopic  dynamics. The crucial point is  that how to connect the 
irreversibility with the time  reversible deterministic microscopic 
dynamics. One outstanding problem is that whether or not the heat 
conduction in one dimensional (1-D) chain obeys the Fourier heat law (normal 
thermal conductivity)? If yes, then 
under what condition. 

The first convincing result of the Fourier heat law in a classical 
system was given by Casati {\it et al} \cite{Casati84}. They studied the 
so-called ding-a-ling model, which is a 1-D chain consists 
of  the fixed equidistant hard-point particle harmonic oscillators, and 
inbetween two fixed particles there is a free particle. The particles 
have the same mass. The two ends of the chain are put into two thermal 
reservoirs. Classically, this system can be changed from  
integrable to fully chaotic by adjusting the system parameter. They found 
that the key ingredient for the normal thermal conductivity 
is chaos. Later on, Prosen and Robnik \cite{PR92} 
have studied the ding-dong model by three different numerical methods and 
verified the Fourier heat law.
The ding-dong model is a modification of the ding-a-ling model. The only 
difference is that in the ding-dong model the fixed harmonic oscillators are 
allowed to collide and there is no free hard-point particles inbetween.
Furthermore, they have studied the temperature dependence of the 
thermal conductivity and found that it increases monotonically with the 
temperature.

Most recently, Lepri {\it et al}\cite{LLP97} have studied the 
Fermi-Pasta-Ulam (FPU) $\beta$ model. This model represents the simplest 
anharmonic approximation of a monoatomic solid. They put the 
Nos$\acute{e}$-Hoover 
thermostats act on the first and the last particle keeping constant 
temperature $T_+$ and $T_-$, respectively. They shown that there exists a 
simple nontrivial scaling relation for the increasing number of 
particles. The thermal conductivity, however, diverges approximately as 
$N^{1/2}$. N is the number of particles. They claimed that chaos is 
not sufficient to ensure the Fourier heat law.

In this paper, we shall investigate the mechanism 
leading to the Fourier heat law. In other words, we would like to answer the 
question: under what condition the heat
conduction of a 1-D many-body Hamiltonian system having the
Fourier heat law. To this end, we shall consider different models, such as
the Frenkel Kontorova (FK) model and the harmonic dissipative model etc..  
We will show that by invoking a simple mechanism we can obtain a rather
satisfactory explanation to all existing numerical results, qualitatively
and quantitatively.  The possible connection with the experimental
results is also discussed. 

{\it Normal thermal conductivity}. --- 
Either the ding-a-ling or the ding-dong model is more or less an 
artificial model.
We would like to turn to a more realistic model, which is close to 
true physical system, i.e. the Frenkel Kontoroval model. It describes a 
particle (atom) chain
connected by harmonic springs subject to an external sinusoidal potential.
It has been widely used to model crystal dislocation, charged
density wave, magnetic spirals and absorbed epitaxial monolayers etc. in
condensed matter physics\cite{Selke92}. This model displays very rich 
interesting phenomena. However, we shall not discuss in detail all this 
properties in this 
paper, for more details please see Refs.\cite{Selke92,Aubry}. Our 
attentions are focused on the thermal conductivity in this paper.  

The existence of the thermal conductivity
of this model has been proved by Gillan and 
Holloway by using different numerical techniques \cite{GH85}.
The classical Hamiltonian of the standard FK model is
\begin{equation}
{\cal H} = \sum_i \frac{P_{i}^{2}}{2m} + 
\frac{\gamma}{2}(X_i - X_{i-1} - a)^2 - 
\frac{A}{(2\pi)^2}\cos\frac{2\pi X_i}{b}. 
\label{Clham1}
\end{equation}
For convenience of numerical calculations, we shall 
scale this Hamiltonian into a  dimensionless one,
\begin{equation}
H = \sum_i \frac{p_{i}^{2}}{2} + 
\frac{1}{2}(x_i - x_{i-1} - \mu)^2 - 
\frac{K}{(2\pi)^2}\cos x_i. 
\label{Clham2}
\end{equation}
By doing this, we have a new effective dimensionless temperature $T$. The 
real temperature $T_r$ is related to $T$ through the following relation,
\begin{equation}
T_r = \frac{m\omega^2_0 b^2}{k_B} T
\label{Temp}
\end{equation}
where $m$ is the mass of the particle,  $\gamma$ the elastic constant, and 
$b$ the period of external potential, which is unit after scaling; $a$ 
the equilibrium distance of the particle, it is $\mu=a/b$ 
after scaling; $K=A/\gamma a^2$ is a rescaled strength of external 
potential. $\omega^2_0 =\gamma/m$ is the oscillator frequency.
$k_B$ the Boltzman constant. In this paper, the winding number in FK 
model is kept at $1/3$.

It is helpful by establishing the above relationship (\ref{Temp}). It 
can give us a very useful information about the corresponding 
true temperature to that one we used and gain some physical insights.
For instance, for the typical values of atoms,
$$
\begin{array}{ll}
b \sim 10^{-10} cm,& \omega_0\sim 10^{13} sec^{-1},\\
m\sim 10^{-26} - 10^{-27} kg,& k_B = 1.38\times 10^{-23} J K^{-1},
\end{array}
$$
we have $T_r \sim 10^2-10^3T$, which means that the room 
temperature corresponds to the dimensionless temperature $T$ about the
order of $0.1-1$. So, if $T$ is very high such as 
to $10^2$, then the actural temperature is about $10^4-10^5$ degree, at
this temperature the displacment of the particle from its equilibrium 
can be up to the order of 10, 
which we think is unrealistic for physical systems. Therefore, like 
Casati {\it et al} \cite{Casati84}, 
we always keep $T$ at very small values in our numerical simulations.

The Hamiltonian (\ref{Clham2}) is a very special 
case.  In fact, we can write it into a general form,
\begin{equation}
H = \sum_i H_i, \qquad H_i=\frac{p_{i}^{2}}{2} + V(x_{i-1},x_i) + 
U(x_i). 
\label{Clham3}
\end{equation}
Here, $V(x_{i-1},x_i)$ stands for the interaction potential of the 
nearest-neighbor particles; $U(x_i)$ is a periodic external potential which 
is an analog of the lattice, and as we shall see later that it plays a 
crucial role in determining the behavior of the thermal conductivity.
If $U(x_i)$ vanishes and $V(x_{i-1},x_i)$ takes the anharmornic form,
Eq. (\ref{Clham3}) is then FPU $\beta$ 
model which has been discussed by Lepri {\it et al} \cite{LLP97}. 
Therefore, a variety of 1-D  models can be put into the 
framwork of Eq.(\ref{Clham3}). 
By changing the form $V(x_{i-1},x_i)$
and $U(x_i)$, we will obtain different thermal conductive 
behaviors.

To study the heat conduction in 1-D model,
we choose the same approach as that used by Lepri {\it et al} \cite{LLP97}, 
namely, 
two Nos$\acute{e}$-Hoover thermostats \cite{NH84} are put on the first and 
last particle, 
keeping the temperature at $T_+$ and $T_-$, respectively. The equations 
of motion of these two particles are determined by,
\begin{equation}
\begin{array}{l}
\ddot{x}_1 = -\zeta_+ \dot{x}_1 + f_1 - f_2,\\
\ddot{x}_N = -\zeta_- \dot{x}_N + f_N - f_{N+1},\\
\dot{\zeta}_+ = \frac{\dot{x}^2_1}{T_+} -1,\qquad
\dot{\zeta}_- = \frac{\dot{x}^2_N}{T_-} -1.
\end{array}
\label{eqm1}
\end{equation}
The equation of motion for the central particles is,
\begin{equation}
\ddot{x}_i = f_i - f_{i+1},\qquad i=2,\dots,N-1,
\label{eqm2}
\end{equation}
where $f_i = -V'(x_{i-1}-x_i)-U'(x_i)$ is the force acting on the 
particle. $x_0=0$ and $x_{N+1} =0$.
\begin{figure}
\epsfxsize=8cm
\epsfbox{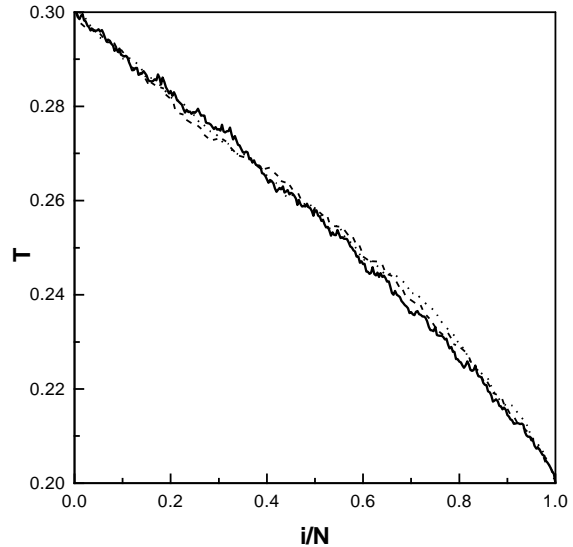}
\vspace{-3.5cm}
\narrowtext
\caption{
Temperature profile for the FK model (\ref{Clham2}) with parameter 
$K=5$. $T_{+}=0.3$, $T_-=0.2$. The average is taken over $ 
10^6$ interval after the transient time $10^4$. The particle numbers are 
300 (solid line), 200 (dashed line) and 100 (dotted line), respectively.
 } 
\end{figure}

We have carried out extensive numerical simulations
with a large 
range of parameters of $N$, $T_{\pm}$ and $K$ for a variety forms of 
$V(x_{i-1},i)$ and $U(x_i)$. We used the seventh-order and eighth-order 
Runge-Kutta algorithm, which provides us much stable and more accurate
results than that from the usual fifth-order Runge-Kutta method.
The spatial temperature profile for standard FK model is shown in Fig. 1. 
It is clear that although our FK model has 
an additional external potential, if its strength $K$ is sufficient 
large (compared with the temperature), we can obtain the same scaling 
relation as that obtained by Lepri {\it et al}. This scaling indicates that 
the temperature gradient scales as $N^{-1}$. We have confirmed that this 
scaling relation is also
true for many different modified FK models. For example, we have changed 
$V(x_{i-1},x_i)$ to the anharmonic case as discussed by Lepri {\it et 
al}, or 
changed the external potential $U(x_i)$ to that one with higher harmonic 
term, such as,
\begin{equation}
 U(x_i)=-\frac{K_1}{(2\pi)^2}\cos(2\pi x_i) - \frac{K_2}
{(4\pi)^2}\cos(4\pi x_i).
\label{U2}
\end{equation}

The derivation of the heat flux of the $i$th particle differs
slightly from that of Lepri {\it et al}. 
The local heat flux $J(x,t)$, which is defined by the 
continuity equation. Taking the 
volume integration on both sides of this equation, we can 
obtain 
\begin{equation}
J_i -J_{i-1} =\dot{x}_i \frac{\partial V}{\partial x_i}(x_i, x_{i+1}) 
-\dot{x}_{i-1} \frac{\partial V}{\partial x_{i-1}}(x_{i-1}, x_i). 
\label{fluxeq2}
\end{equation}
Thus the heat flux is defined by
\begin{equation}
J_i =\dot{x}_i \frac{\partial V}{\partial x_i}(x_i, x_{i+1}).
\label{fluxdef}
\end{equation}
Numerically, the time average 
$J = \langle J_i(t)\rangle$ is independent of the index $i$ for long 
enough time.
\begin{figure} 
\epsfxsize=8cm 
\epsfbox{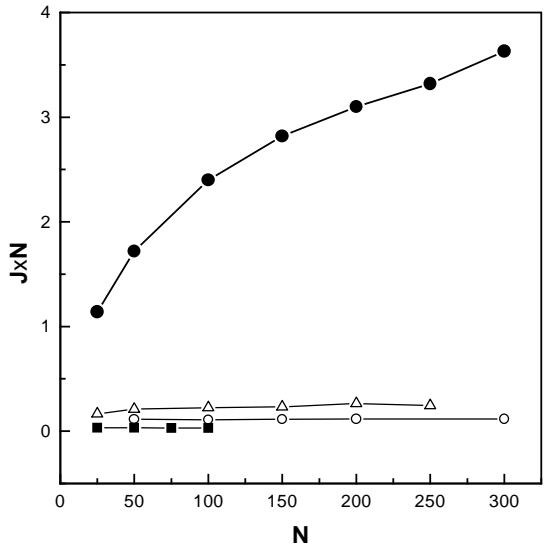} 
\vspace{-3.5cm}
\narrowtext 
\caption{ 
$J\times N$ versus the number of particles $N$ for
different models. $T_+=0.3$ and $T_-=0.2$ for all cases. The solid circle 
represents the results of the FPU
$\beta$ model ($\beta=0.5$).
Open circle is the results of FK model given by Eq. (\ref{Clham2}) with
$K=5$; solid square for FK model with an external potential (\ref{U2}), 
with $K_1=5$ and $K_2=15$; and the open triangle for FK model with $V = 
x^2/2 + \beta 
x^4/4$, $U=-K \cos(2\pi x)/(2\pi)^2$, $\beta=0.9, K=5$. The lines are 
draw to guide the eyes.
}
\end{figure}

The N dependence of $J\times N$  is plotted in Fig. 2 for different 
models. As is easily seen that for different FK models (with different 
$V(x_{i-1},x_i)$ and/or different $U(x_i)$), as long as $U(x_i)$ is 
nonzero and at sufficient lower temperature, $J\times N$ is a constant 
implying $1/J$ diverges with $N$. Since 
the temperature gradient vanishes as $N^{-1}$ as is shown in Fig. 1, thus
the Fourier heat law is justified. 

{\it Abnormal thermal conductivity.} ---
Things become very different, if $U(x_i)$ vanishes. In this case, 
the heat conduction does not obey the Fourier heat law 
neither for the harmonic form $V(x_{i-1},x_i)$ 
nor for the anharmonic form 
such as the FPU $\beta$ model discussed by Lepri {\it et al}. 
Our results for FPU $\beta$ model at very low temperature shown in Fig. 
3 also
demonstrate that $J\times N$ diverges as approximately $N^{1/2}$ which 
means that the thermal conductivity diverges as $N^{1/2}$. This agrees 
with that of Lepri {\it et al} at much higher temperature.

Based on the above results, we are convinced to conclude that 
the key point of the normal thermal 
conductivity is the {\em periodic} external potential, which is analogous to 
the lattice. 

If the lattice is absent, and the interparticle potential
is harmonic, then no phonon-phonon 
interaction exists, thus the heat transfer would take place at the
speed of sound and the thermal conductivity would be infinite as was
pointed out by Debye in 1914. 
(However, if we add a dissipative term to the harmonic oscillator chain, 
then we could obtain the Fourier heat law, even though we have not lattice. 
This is because the dissipation, the heat radiates during the transport. 
Our numerical results have verified this. But we will discuss this more 
in detail in another paper \cite{hlzz97}.)

In the case of having an anharmonic interparticle potential 
$V(x_{i-1},x_i)$ such as that in the FPU model, the phonon-phonon 
interaction is produced due to the anharmonicity. Although the temperature 
gradient can be formed, nevertheless, as is shown by the work of Lepri {\it 
et al} at high 
temperature as well as ours at low temperature, the thermal conductivity 
diverges. 

As long as the lattice exists, 
the phonons will be scattered by it and results in the thermal 
resistance, eventually leads to the Fourier heat 
law. In the ding-a-ling model and the ding-dong models the fixed harmonic 
oscillator plays the role of the lattice, whereas in the FK model, it is the 
periodic external potential. In these three cases
the Fourier heat law is justified numerically. Thus we believe that it 
might be a general rule that if the phonon-lattice interaction is dominant, 
the heat conduction will obey the Fourier heat law, no matter whether the 
interparticle interaction is harmonic or anharmonic.

{\it Temperature dependence of $J$}. --- As discussed above, the crucial 
point of the Fourier heat law is the
phonon-lattice interaction. 
The mean free path  of the phonons is determined by the density of 
lattice and does not change with the temperature. By increasing  
the temperature, more and more high energy phonons are excited, which 
results in the growth of the heat flux, thus the increment of the 
thermal conductivity. Whereas in the absence of
lattice, increase  temperature will produce
more phonons, which in turn reduce the phonons' mean free path, 
consequently 
decreases the heat flux. Therefore, the temperature dependence behaviour 
for normal and abnormal thermal conductivity should be very different.
Our numerical calculations exactly demonstrate this point.

In Fig. 3, we plot the temperature dependence of heat flux 
for different models. The particle number is kept at 
$N=100$, and in all cases the temperature 
difference is fixed at
$\Delta T= T_+ - T_-=0.1$, thus $J$ has the same behavior of the thermal 
conductivity $\kappa$.
 For the FPU 
$\beta$ model ($\beta=0.9$, solid circle), the heat flux decreases 
monotonically with 
temperature, whereas in the standard FK model with ($K=5$, solid triangle) 
increases with temperature. 

Another very important thing deserve noting is the case in which the 
anharmonicity and the external potential coexists.
It seems that this case is closer to the real 
physical system than 
others. We have performed the numerical simulation by using $V=x^2/2 + 
\beta x^4/4$ and  $U=-K\cos(2\pi x)/(2\pi)^2$ in Eq.(\ref{Clham3}). The 
temperature dependence of $J$ is shown in Fig. 
3 (solid square). The heat flow is affected not only by 
the phonon-lattice interaction, but also by the 
phonon-phonon interaction. 
At low temperature region, the factor 
determining the heat conduction is the phonon-lattice interaction, 
therefore, the heat conduction obeys the Fourier heat law, but the heat 
flux is bigger than that case of the standard FK model (solid triangle) 
due to the anharmonicty which produces more phonons to transfer heat. 
The anharmonicity becomes more and more important when the 
temperature is increased,  this is why at higher 
temperature region, a relative flat region shows up in Fig. 3. 
Furthermore, it must be noted that our numerical results shown that for 
FK model shown in Fig. 3 the Fourier heat law is valid only at lower 
temperature region $T<1$, at higher temperature the Fourier heat law 
broken down due to the reason mentioned before.
\begin{figure}
\epsfxsize=8cm
\epsfbox{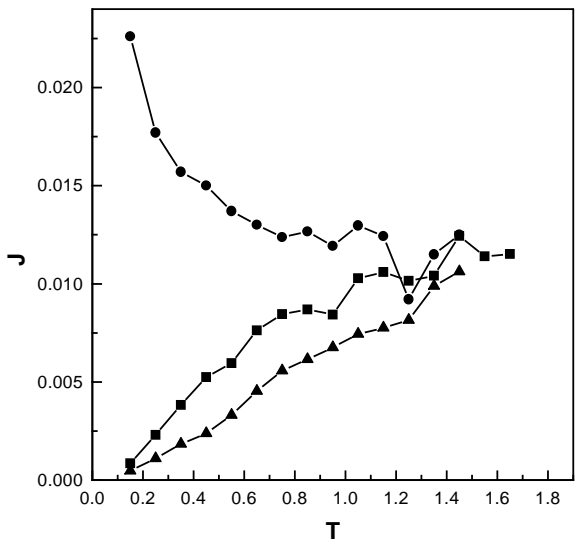}
\vspace{-3.5cm}
\narrowtext
\caption{
The temperature dependence of heat flux $J$ for the FPU $\beta$ model 
($\beta=0.9$, 
solid circle), the standard FK model Eq.(\ref{Clham2}) (K=5, solid 
triangle), and the FK models with an anharmonic 
interparticle potential $V(x) = x^2/2+\beta x^4/4$ and external potential 
$K\cos(2\pi x)/(2\pi)^2$ at $\beta=0.9, K=5$ (solid square). The line is 
draw just for guiding the eyes. 
} 
\end{figure}
From many experimental results (see e.g. 
the book of Srivastava\cite{Sri90} or other
textbooks of solid state physics, such as that of Kittel\cite{Kittel}), 
we observe that the thermal 
conductivity increases with temperature at lower temperature region,
whereas  it decreases at high temperature. 
This can be understood well from the mechanism discussed in this 
paper.

In summary, by studying the dynamical equations of the 1-D particle 
chain, we understood more about the heat conduction mechanism.
Our numerical results as well as others up to date 
confirm our conjecture that the phonon-lattice interaction is the 
key factor for the Fourier heat law. Only the phonon-phonon 
interaction cannot give rise to the Fourier heat law, instead we will have 
the 
abnormal thermal conductivity, i.e. the thermal conductivity diverges 
as the particle's number. In the former case, the thermal conductivity 
grows with the temperature monotonically, whereas in the latter case it 
decreases.

We would like to thank Dr. Zhigang Zheng for many useful and stimulating 
discussions and bringing our attention  to the reference \cite{GH85}; 
Thanks also go to Dr. Jilin Zhou for kindly providing the seventh-order and 
eighth-order 
Runge-Kutta integration program. This work is supported in part by Hong 
Kong Research Grant
Council and the Hong Kong Baptist University Faculty
Research Grant.

\end{multicols}
\end{document}